\documentclass[twocolumn,showpacs,aps,pre]{revtex4}
\usepackage{amsmath}
\usepackage{amssymb}
\usepackage{graphicx}
\usepackage{graphics}

\begin{document}

\title{Polymer drift in a solvent by force acting on one polymer end}
\author{Semjon Stepanow$^1$ and Norio Kikuchi$^{1,2}$}
\affiliation{$^1$ Institut f\"{u}r Physik, Martin-Luther-Universit\"{a}t
Halle-Wittenberg, D-06099 Halle, Germany\\
$^2$ Centre for Condensed Matter Theory, Department of Physics, Indian
Institute of Science, Bangalore 560 012 India }
\date{\today}

\begin{abstract}
We investigate the effect of hydrodynamic interactions on the non-equilibrium
drift dynamics of an ideal flexible polymer pulled by a constant force applied at one end of the polymer using the perturbation theory and
the renormalization group method.
For moderate force, if the polymer elongation is small, the hydrodynamic interactions are not screened and the velocity and the longitudinal elongation of the polymer are computed using the renormalization group method. Both the velocity and elongation are nonlinear functions of the driving force in this regime.
For large elongation we found two regimes. For large force but finite chain length $L$ the hydrodynamic interactions are screened.
For large chain lengths and a finite force the hydrodynamic interactions are only partially screened, which in three dimensions results in unusual logarithmic corrections to the velocity and the longitudinal elongation.
\end{abstract}

\maketitle


\section{Introduction}

\label{intro}

Understanding the dynamics of polymers in solutions is of great importance
in many bio-chemical and industrial process. In dilute solution,
hydrodynamic interaction between different parts of the polymer chain plays
a crucial role in the determination of dynamic properties \cite{doi-edwards}. In the classical theories which go back to Kirkwood and Zimm \cite{kirkwood,zimm}, the main focus of research was to study the quantities such as the diffusion
constant, the dynamic intrinsic viscosity, and the dynamic structure factor
of the polymer. Recent advances in experimental techniques make it possible
to explore the motion of individual polymer under hydrodynamic flow, thermal
noise, or external fields.
For instance, Chu et al. \cite{larson-chu97,perkins-smith-chu97,schoeder-chu03} experimentally studied the behavior of polymers in
different flow conditions with the emphasis on biological applications. See also related theoretical studies {\cite{brochard93}-\cite{zimmermann-epje00}.

In this paper we study drift of an ideal
flexible polymer pulled by a constant driving force applied at one polymer end. Classical theory of dragging objects dates back to Einstein's
Brownian motion and Stokes' theory in the hydrodynamic
medium, which yields the velocity of the dragged object, $\mathrm{v}_B = F/f_0$
with $f_0 =C{\eta}_sR$.
Unlike the drag of these simple objects, the dynamics of the polymer pulled at polymer end is more
complicated due to the interplay between the internal degrees of freedom of polymer, hydrodynamic interactions, and inhomogeneity of chain configurations, which is due to the pulling force acting at polymer end.
The drift by a force applied at polymer end is different from the case, if the external force is homogeneous and acts on all monomers of the chain.

In neglecting the hydrodynamic interactions i.e. in the Rouse theory
the average velocity and the chain elongation are given by expressions
\begin{equation}
\mathrm{v}^{z}\simeq \frac{F}{f_{0}N},  \label{v-R}
\end{equation}%
\begin{equation}
\left\langle r^{z}(0,t)-r^{z}(L,t)\right\rangle =\frac{FLl}{2dk_{B}T},
\label{r-R}
\end{equation}
where $d$ is the dimensionality, $F$ is the pulling force, and $f_0$ denotes the monomer friction coefficient. For small forces the polymer is expected to have the shape of coil with the consequence that the hydrodynamic interaction is not screened (Zimm regime), and the velocity is described by Stokes formula
\begin{equation}
\mathrm{v}^{z}\simeq \frac{1}{6\pi \eta _{s}R_{h}}F  \label{v-Z}
\end{equation}
with $R_{h}\simeq l\sqrt{N}$ being the hydrodynamic radius of the polymer, and $\eta _{s}$ is the solvent viscosity. The Zimm behavior (\ref{v-Z}) is described in the framework of the renormalization group method as due to the renormalization of the friction coefficient, which becomes scale dependent in the regime of strong hydrodynamic interaction \cite{freed}-\cite{stepanow}.

For moderate forces a crossover between (\ref{v-R}) and (\ref{v-Z}) as well as a nonlinear dependence on the pulling force is expected to apply. For large forces the effect of the hydrodynamic interaction is expected to be weak due to polymer stretching, so that a Rouse-type behavior should be observed.

In the present work we calculate the first-order corrections to the velocity and the longitudinal elongation. The renormalization group method allows us to establish the behavior of the quantities under consideration for moderate forces. We also show that in $d=3$ the first-order corrections
to the velocity and the longitudinal size of the polymer in powers of the
hydrodynamic interactions logarithmically depend on the parameter $\beta F\sqrt{lL/2d}$, which demonstrates that hydrodynamic
interactions are only partially screened for long polymer pulled at finite force.

The paper is organized as follows. After an introduction to the formalism in section \ref{sect1}, we perform the perturbational computation of the average velocity
in section \ref{numer} and the longitudinal polymer size under the drift in
section \ref{perturb_longitudinalsize}. In section \ref{results}, the
velocity and the longitudinal size are analyzed for small and large force
using the renormalization group analysis. The work is summarized in section \ref{concl}.

\section{The formalism}

\label{sect1}

The drift dynamics of a polymer chain is described by the Kirkwood
diffusion equation, which in the
continuum limit is given by \cite{doi-edwards}%
\begin{eqnarray}
&&\frac{\partial P\left( \left\{ \mathbf{r}(s)\right\} ,t\right) }{\partial t%
}=\int_{0}^{L}ds_{2}\int_{0}^{L}ds_{1}\frac{\delta }{\delta \mathbf{r}(s_{1})%
} \times \notag \\
&&\left[ D_{0}\delta ^{\mu \nu }\delta (s_{1}-s_{2})+T^{\mu \nu }(\mathbf{r}%
(s_{1})-\mathbf{r}(s_{2}))\right] \times \notag \\
&&\left[ \frac{\delta H_{0}}{\delta \mathbf{r}(s_{2})}-\beta F_{ext}(\mathbf{%
r}(s_{2}))+\frac{\delta }{\delta \mathbf{r}(s_{2})}\right] P.  \label{eq1}
\end{eqnarray}%
The Oseen tensor (multiplied by $k_{B}T$) in $d$-dimensions is
defined through its Fourier transform as%
\begin{eqnarray*}
&&T^{\mu \nu }(\mathbf{r}(s_{1})-\mathbf{r}(s_{2}))= \\
&&\frac{k_{B}T}{\eta _{s}}\int \frac{d^{d}q}{(2\pi )^{d}}\frac{1}{q^2}%
(\delta ^{\mu \nu }-\frac{q^{\mu }q^{\nu }}{q^{2}})e^{i\mathbf{q}(\mathbf{r}%
(s_{1})-\mathbf{r}(s_{2}))}.
\end{eqnarray*}%
The free energy (multiplied by $k_{B}T$) and the pulling force applied in $z$%
-direction are given by%
\begin{equation}
H_{0}=\frac{d}{2l}\int_{0}^{L}ds\left( \frac{d\mathbf{r}(s)}{ds}\right)
^{2},\ \ F_{ext}^{\mu }(\mathbf{r}(s))=\delta ^{\mu z}F\delta (s).
\label{eq3}
\end{equation}%
The Rouse modes $\boldsymbol{\xi }_{k}$ are the Fourier coefficients in the
expansion of the position vector $\mathbf{r}(s)$ ($0\leq s\leq L$) according
to $\mathbf{r}(s)=\sum\limits_{k=0}^{\infty }Q_{sk}\boldsymbol{\xi }_{k}$
where the basis functions 
\begin{equation}
Q_{sk}=\left\{
\begin{array}{cc}
\frac{1}{\sqrt{L}}, & k=0 \\
\sqrt{\frac{2}{L}}\cos \frac{\pi sk}{L}, & k=1,2,3,\cdots%
\end{array}%
\right.  \label{eq4}
\end{equation}%
are orthogonal and complete in the interval ($0$, $L$).

The Kirkwood diffusion equation can be rewritten in terms of the normal
coordinates%
\begin{equation}
\frac{\partial P(\xi ,t;\xi ^{0},t_{0})}{\partial t}=L_{0}P+L_{i}P,
\label{eq5}
\end{equation}%
where $\xi $ is abbreviation for $\xi _{k}$, and the operators $L_{0}$ and $%
L_{i}$ are given by
\begin{equation}
L_{0}=D_{0}\nabla _{k}^{2}+D_{0}\lambda _{k}\nabla _{k}^{\mu }\xi _{k}^{\mu
}-D_{0}\beta F^{\mu }Q_{0k}\nabla _{k}^{\mu },  \label{eq6}
\end{equation}%
\begin{equation}
L_{i}=\nabla _{k}^{\mu }T_{km}^{\mu \nu }\left( \nabla _{m}^{\nu }+\lambda
_{m}\xi _{m}^{\nu }\right) -\nabla _{k}^{\mu }\tilde{T}_{k0}^{\mu \nu }\beta
F^{\nu }  \label{eq7}
\end{equation}%
with abbreviations $\nabla _{k}{\equiv} {\delta }/{\delta
{\xi}_{k}}$,
\begin{eqnarray*}
T_{km}^{\mu \nu } &=&\int_{0}^{L}ds_{2}\int_{0}^{L}ds_{1}Q_{s_{1}k}T^{\mu
\nu }(\mathbf{r}(s_{1})-\mathbf{r}(s_{2}))Q_{s_{2}m}, \\
\tilde{T}_{k0}^{\mu \nu } &=&\int_{0}^{L}ds_{1}Q_{s_{1}k}T^{\mu \nu }(%
\mathbf{r}(s_{1})-\mathbf{r}(0)),
\end{eqnarray*}%
and $r^{\mu }(s_{1})-r^{\mu }(s_{2})=\sum\limits_{n=0}^{\infty
}(Q_{s_{1}n}-Q_{s_{2}n})\xi _{n}^{\mu }$.

In the following we consider the hydrodynamic interaction as perturbation,
and use the Rouse chain pulled by a force at one end as a reference state.
The\ solution of Eq.~(\ref{eq5}) with $L_{i}=0$ gives the transition
probability density of the reference Rouse state 
\begin{eqnarray}
&&P_{0}(\xi ,t;\xi ^{0},t_0)=\prod\limits_{k\neq 0}\frac{1}{\left( 4\pi
f_{k}(t-t_0)\right) ^{d/2}} \times \notag \\
&&\exp \left( -\frac{\left( \xi _{k}^{\mu }-\beta F^{\mu }Q_{0k}/\lambda
_{k}-\xi _{k}^{0\mu }a_{k}(t-t_0)\right) ^{\!2}}{4f_{k}(t-t_0)}\right) \times
\notag \\
&&\frac{1}{\left( 4\pi D_{0}t\right) ^{d/2}}\exp\! \left(\! -\frac{\left(
\xi _{0}^{\mu }-D_{0}\beta F^{\mu }Q_{00}t-\xi _{0}^{0\mu }\!\right) ^{\!2}}{%
4D_{0}t}\right)  \label{eq8}
\end{eqnarray}%
%
%
%
%
%
%
with%
\begin{equation*}
a_{k}(t)=\exp (-D_{0}\lambda _{k}t),\ \ \ f_{k}(t)=\frac{1-\exp
(-2D_{0}\lambda _{k}t)}{2\lambda _{k}},\
\end{equation*}%
where $D_{0}=k_{B}Tl/f_0$ is the monomer self-diffusion coefficient, and $\
\lambda _{k}=\frac{d}{l}\left( \frac{\pi k}{L}\right) ^{2}$ are the Rouse
eigenvalues.

To perform perturbational computations of the velocity, the longitudinal
size, etc. of the polymer chain under the pulling force, we rewrite the
Kirkwood diffusion equation as an integral equation%
\begin{eqnarray}
&&P(\xi ,t;\xi ^{0},t_{0})=P_{0}(\xi ,t;\xi
^{0},t_{0})+\int_{t_{0}}^{t}dt^{\prime }\times   \notag \\
&&\int D\xi ^{\prime }P_{0}(\xi ,t;\xi ^{\prime },t^{\prime })L_{i}(\xi
^{\prime })P(\xi ^{\prime },t^{\prime };\xi ^{0},t_{0}).  \label{eq9}
\end{eqnarray}%
Iteration of (\ref{eq9}) results in the perturbation expansions in powers of
the hydrodynamic interaction, which is symbolically written: $%
P=P_{0}+P_{0}L_{i}P=P_{0}+P_{0}L_{i}P_{0}+P_{0}L_{i}P_{0}L_{i}P=\cdots $.

\section{Perturbational computation of the velocity}

\label{numer} Our aim is now to calculate the velocity of the polymer chain
up to the first order in hydrodynamic interaction using Eq.~(\ref{eq9}). The
center of mass $\mathbf{r}_{c}$ of the polymer chain is expressed through
the center of mass Rouse mode $\boldsymbol{\xi }_{k=0}$ according to $%
\mathbf{r}_{c}=\boldsymbol{\xi }_{k=0}/\sqrt{L}=(1/L)\!\int_{0}^{L}\!\!ds\,%
\boldsymbol{r }(s)$. Hence, the computation of the average velocity reduces
to the computation of the expectation value of the zero mode $\left\langle %
\boldsymbol{r}_{c}(t)\right\rangle =\left\langle \boldsymbol{v}%
_{c}(t)\right\rangle {t}$:%
\begin{equation}
\left\langle \boldsymbol{\xi }_{k=0}(t)\right\rangle \equiv \int D\xi %
\boldsymbol{\xi }_{k=0}P(\xi ,t;\xi ^{0},t_{0}).  \label{vel_0}
\end{equation}%
The computation of (\ref{vel_0}) to the zero order in powers of the
hydrodynamic interaction (i.e. in the Rouse model) using the unperturbed
distribution function (\ref{eq8}) yields in the steady state%
\begin{equation*}
\left\langle \xi _{k=0}^{z}(t)\right\rangle _{0}=\frac{lF}{f\sqrt{L}}t,
\end{equation*}%
where we take the force, which is applied at $s=0$, to be directed along the $z$%
-axes, $F=F^{z}$. For the velocity we obtain the expression
$\left\langle \boldsymbol{v}_{c}^{z}(t)\right\rangle _{0}=F/(fN)$, which is in agreement with Rouse theory. The
first-order correction is derived from Eq.~(\ref{eq9}) by replacing $P$ on
the right-hand side by $P_{0}$ as
\begin{eqnarray}
&&\left\langle \xi _{k=0}^{z}(t)\right\rangle
_{1}=\int_{t_{0}}^{t}dt^{\prime }\int_{0}^{L}ds_{2}\int_{0}^{L}ds_{1}\int
\frac{d^{d}q}{(2\pi )^{d}}Q_{s_{2}0}\times   \notag \\
&&T^{zz}(q)Q_{s_{1}0}\int D\xi ^{\prime }e^{i\mathbf{q}\sum\limits_{n=0}^{%
\infty }(Q_{s_{1}n}-Q_{s_{2}n})\boldsymbol{\xi }_{n}^{\prime }}\beta
FQ_{00}\times   \notag \\
&&P_{0}(\xi _{k}^{\prime })P_{0}(\xi _{0}^{\prime }-\xi _{0}^{0}-D_{0}\beta
FQ_{00}(t^{\prime }-t_{0}))  \label{eq10}
\end{eqnarray}%
where $T^{{\mu }{\nu }}(q)=\frac{k_{B}T}{{\eta }_{s}}\frac{1}{q^{2}}(\delta
^{\mu \nu }-\frac{q^{\mu }q^{\nu }}{q^{2}})$. The integrations over $\xi
^{\prime }$ yields%
\begin{eqnarray}
&&\left\langle \xi _{k=0}^{z}(t)\right\rangle _{1}=\beta
FQ_{00}\int_{t_{0}}^{t}dt^{\prime }\int_{0}^{L}ds_{2}\int_{0}^{L}ds_{1}
 \times \notag \\
&&\int \frac{d^{d}q}{(2\pi )^{d}}Q_{s_{2}0}T^{zz}(q)Q_{s_{1}0}\times   \notag
\\
&&\exp \left( iq^{z}F\beta \sum\limits_{n=1}^{\infty }\frac{%
Q_{s_{1}n}-Q_{s_{2}n}}{\lambda _{n}}Q_{0n}-\right.   \notag \\
&&\left. q^{2}\sum\limits_{n=1}^{\infty }\frac{(Q_{s_{1}n}-Q_{s_{2}n})^{2}}{%
2\lambda _{n}}\right) .  \label{eq11}
\end{eqnarray}%
The evaluation of the sums in Eq.~(\ref{eq11}) gives%
\begin{eqnarray}
&&\left\langle \xi _{k=0}^{z}(t)\right\rangle
_{1}=\int_{t_{0}}^{t}dt^{\prime }\int_{0}^{L}ds_{2}\int_{0}^{L}ds_{1}\int
\frac{d^{d}q}{(2\pi )^{d}}Q_{s_{2}0}\times   \notag \\
&&T^{zz}(q)Q_{s_{1}0}\beta FQ_{00}\exp \left( iq_{z}F_{z}{\beta }%
b(s_{1},s_{2})-aq^{2}\right)   \label{eq12}
\end{eqnarray}%
where the quantities $a$ and $b$ are defined as follows
\begin{eqnarray*}
a &\equiv &\sum\limits_{n=1}^{\infty }\frac{(Q_{s_{1}n}-Q_{s_{2}n})^{2}}{%
2\lambda _{n}}=\frac{l}{2d}\left\vert s_{1}-s_{2}\right\vert ,\ \  \\
b &\equiv &\sum\limits_{n=1}^{\infty }\frac{Q_{s_{1}n}-Q_{s_{2}n}}{\lambda
_{n}}Q_{0n}=\frac{l}{d}(s_{2}-s_{1})(1-\frac{s_{1}+s_{2}}{2L}).
\end{eqnarray*}%
Performing the straightforward integration over $q$ we finally obtain

\begin{eqnarray}
&&\mathrm{v}_{c}^{z}(t)=\frac{\left\langle \xi _{k=0}^{z}(t)\right\rangle }{%
\sqrt{L}t}=\frac{F}{f_{0}N}(1+  \notag \\
&&\frac{1}{2^{2}}\frac{f_{0}}{d\eta _{s}}\left( \frac{d}{2\pi l}\right)
^{d/2}L^{2-d/2}\times  \notag \\
&&\int_{0}^{1}dx_{2}\int_{0}^{x_{2}}dx_{1}\frac{A(y)}{(x_{2}-x_{1})^{d/2-1}}%
+\cdots)  \label{eq13}
\end{eqnarray}%
where the function $A(y)$ and its argument $y$ are respectively given by%
\begin{eqnarray}
A(y) &=&\frac{(d-1)(d-2)}{y^{4}}e^{-y^{2}}-\frac{(d-1)\left(
d-2-2y^{2}\right) }{y^{d}}\times  \notag \\
&&\left( \frac{4-d}{2}\Gamma \left( \frac{d-4}{2},y^{2}\right) +\Gamma
\left( \frac{d-2}{2}\right) \right) ,  \label{eq14}
\end{eqnarray}%
\begin{equation}
y=\beta F\sqrt{\frac{lL}{2d}}(x_{2}-x_{1})^{1/2}(1-\frac{x_{2}+x_{1}}{2}). \label{def-y}
\end{equation}%
$\Gamma\!\left(a,z\right)=\int_z^{\infty}\!\!dt\,t^{a-1}e^{-t}$ is the
incomplete gamma function. The function $A(y)$ behaves for small and large
arguments as%
\begin{equation}
A(y)\simeq \left\{
\begin{array}{ll}
\frac{8(d-1)}{d(d-2)}\!-\!\frac{8\left( d-1\right) }{d(d+2)}y^{2}\!+\!\frac{%
4\left( d-1\right) }{(d+2)(d+4)}y^{4}\!+\!\cdots\!, & y\ll 1, \\
\frac{(d-1)(2y^{2}+2-d)}{y^{d}}\Gamma \left( \frac{d-2}{2}\right)\!+\!\frac{%
2(d-1)}{y^{4}}e^{-y^{2}}\!\!+\!\cdots\!, & y\gg 1.%
\end{array}%
\right.  \label{eq15}
\end{equation}%
Integrations over $x_{1}$ and $x_{2}$ in Eq.~(\ref{eq13}), which can be
carried out in the limit $F\rightarrow 0$, yield%
\begin{eqnarray}
&&\int_0^1\!\!dx_2\int_0^{x_2}\!\!dx_1\frac{A(0)}{(x_2-x_1)^{d/2-1}}  \notag
\\
&=&\frac{4}{(6-d)(4-d)}A(0)\overset{d\sim 4}{\rightarrow }\frac{2}{4-d}A(0).
\end{eqnarray}%
Note that the behavior of the first-order correction to the velocity in the
vicinity of four dimensions, which is given by the last expression, plays an
important role in the renormalization group analysis of the velocity.


\section{Perturbational computation of the longitudinal size of the polymer}

\label{perturb_longitudinalsize}

In this section we calculate the longitudinal size of the polymer up to the
first order in hydrodynamic interaction. The longitudinal size of the
polymer is expressed through the Rouse modes according to
\begin{equation}
\left\langle r^{z}(0,t)-r^{z}(L,t)\right\rangle \!=\!\sqrt{\frac{2}{L}}%
\sum_{k=1}^{\infty }\left( 1\!-\!(-1)^{k}\right) \!\left\langle \xi
_{k}^{z}(t)\right\rangle \!.  \label{eq16}
\end{equation}%
The computation to the zero order in hydrodynamic interaction (i.e. in the
Rouse model) in the steady state yields
\begin{equation*}
\left\langle r^{z}(0,t)-r^{z}(L,t)\right\rangle _{\!0}=\frac{FLl}{2dk_{B}T}.
\end{equation*}%
The first-order correction to $\left\langle \xi _{k}^{z}(t)\right\rangle $
is obtained using Eq.~(\ref{eq9}) as
\begin{eqnarray}
&&\left\langle \xi _{k}^{z}(t)\right\rangle _{1}=\beta
F\int_{t_{0}}^{t}dt^{\prime }a_{k}(t-t^{\prime
})\int_{0}^{L}ds_{2}\int_{0}^{L}ds_{1}\times   \notag \\
&&\int \frac{d^{d}q}{(2\pi )^{d}}Q_{s_{2}k}T^{zz}(q)Q_{s_{1}m}Q_{0m}\mid
_{m=0}\times   \notag \\
&&\exp \left( iq^{z}F\beta \sum\limits_{n=1}^{\infty }\frac{%
Q_{s_{1}n}-Q_{s_{2}n}}{\lambda _{n}}Q_{0n}-\right.   \notag \\
&&\left. q^{2}\sum\limits_{n=1}^{\infty }\frac{(Q_{s_{1}n}-Q_{s_{2}n})^{2}}{%
2\lambda _{n}}\right) .  \label{eq17}
\end{eqnarray}%
The integration over $q$ and summations over $k$ and $n$ are performed
similar to those in the preceding section. The longitudinal elongation reads
\begin{eqnarray}
&&\left\langle r^{z}(0,t)-r^{z}(L,t)\right\rangle =\frac{FLl}{2dk_{B}T}(1-%
\frac{1}{2^{2}}\frac{f_{0}}{d\eta _{s}}\times   \notag \\
&&\left( \frac{d}{2\pi l}\right) ^{d/2}L^{2-d/2}B_{r}(\beta F\sqrt{\frac{lL}{%
2d}})+\cdots )  \label{eq18}
\end{eqnarray}%
with
\begin{equation}
B_{r}(\beta F\sqrt{lL/2d})=\int_{0}^{1}dx_{2}\int_{0}^{x_{2}}dx_{1}\frac{%
(2x_{2}-1)A(y)}{(x_{2}-x_{1})^{d/2-1}}. \label{eq18a}
\end{equation}%
The computation of integrals in Eq.~(\ref{eq18a}) over $x_{1}$ and $x_{2}$
for a small force yields%
\begin{equation*}
B_{r}(F\!\rightarrow \!0)=\frac{4}{(8-d)(6-d)}A(0).
\end{equation*}%
The finiteness of the latter at four dimensions means in the context of the
renormalization group method that the parameter $F$, $L$, $l$, and $T$
appearing in the prefactor of Eq.~(\ref{eq18}) do not renormalize. The only
quantity which renormalize is the monomer friction coefficient. The
hydrodynamic interaction results in decreasing the size of the polymer.


\section{Results}

\label{results}

Our aim is now to study the behavior of the velocity and the longitudinal
size of the polymer chain as a function of the driving force $F$ and the chain
length $L$.

\subsection{Small elongation}

The first-order perturbational correction to the velocity is the starting
point to perform the renormalization group (RG) analysis, which enables one
to take into account the effect of hydrodynamic interaction beyond the
first-order. The basic observation to apply RG is that the integral in (\ref%
{eq13}) diverges logarithmically in four dimensions (i.e. the critical dimension
is four; for $d>4$ the hydrodynamic interactions become irrelevant). This
divergence manifests itself in $d<4$ dimensions as $1/(4-d)$ pole. To
regularize the theory these poles in perturbation expansions have to be
removed by an appropriate renormalization of the friction coefficient. In
the limit of a small pulling force the renormalized friction coefficient is
derived from (\ref{eq13}) as
\begin{equation}
f=f_{0}(1-\frac{3}{2\varepsilon }\xi _{0}(L^{\varepsilon /2}-\lambda
^{\varepsilon /2})+\cdots )  \label{eq19}
\end{equation}%
where $\xi _{0}=(f_{0}/(\eta _{s}d))(d/(2\pi l))^{d/2}$ is the expansion
parameter of perturbation expansions in powers of the hydrodynamic
interactions, and $\varepsilon =4-d$. The ultraviolet cutoff $\lambda $ in (%
\ref{eq19}) is introduced to enable the limit to four dimensions. The cutoff
excludes the hydrodynamic interactions between monomers separated along the
chain by the contour length less than $\lambda $, $\left\vert
s_{1}-s_{2}\right\vert <\lambda $. The renormalization of the friction
coefficient (\ref{eq19}) obtained from Eq.~(\ref{eq13}) coincides with that
obtained by studying different problems in polymer dynamics \cite{freed} -%
\cite{stepanow}. Note that in the absence of the excluded volume
interaction, only friction coefficient renormalizes.

In addition to the renormalization of the friction coefficient one should
consider the renormalization of the coupling constant (i.e. expansion
parameter) controlling the strength of the hydrodynamic interaction. The
inspection of Eq.~(\ref{eq13}) or Eq.~(\ref{eq18}) yields the bare
dimensionless expansion parameter as%
\begin{equation}
w_{0}=\frac{f_{0}}{\eta _{s}d}(\frac{d}{2\pi l})^{d/2}L^{\varepsilon /2}.
\label{eq20}
\end{equation}%
It follows from Eq.~(\ref{eq20}) that the coupling constant renormalizes in the same way as the friction coefficient.

The renormalization group is based on the observation that the
regularization, i.e. the elimination of the $1/\epsilon$-poles from perturbation expansions can be performed step by step by changing the
cutoff $\lambda \rightarrow \lambda ^{\prime }\rightarrow \lambda ^{\prime
\prime }\rightarrow \cdots \rightarrow \lambda _{m}$. The
renormalization of the friction coefficient and the strength of the
hydrodynamic interaction due to an infinitesimal change of the cutoff are
given by differential equations. To the one-loop order one obtains
\begin{equation}
\lambda ^{\prime }\frac{\partial f}{\partial \lambda ^{\prime }}=\frac{3}{4}%
w,  \label{eq21}
\end{equation}%
\begin{equation}
\lambda ^{\prime }\frac{\partial w}{\partial \lambda ^{\prime }}=\frac{%
\varepsilon }{2}w-\frac{3}{4}w^{2}+...\equiv \beta (w)  \label{eq22}
\end{equation}%
with the dimensionless effective coupling constant $w=(f/(\eta
_{s}d))(d/(2\pi l))^{d/2}\lambda ^{\prime \varepsilon /2}$. The solutions of
Eqs.~(\ref{eq21}) and (\ref{eq22}) read%
\begin{equation}
f=\frac{f_{0}}{1+\frac{3}{2\varepsilon }\xi _{0}(\lambda _{m}^{\varepsilon
/2}-\lambda ^{\varepsilon /2})},  \label{eq23}
\end{equation}%
\begin{equation}
w=\frac{f}{\eta _{s}d}(\frac{d}{2\pi l})^{d/2}\lambda _{m}^{\varepsilon /2}.
\label{eq24}
\end{equation}%
It follows from Eq.~(\ref{eq23}) and (\ref{eq24}) that at large $\lambda _{m}$, $w$
approaches the fixed-point value $w^{\ast }=2\varepsilon /3$, which
corresponds to the zero point of the Gell-Mann - Low function $\beta (w)$.
At the fixed-point the effective friction coefficient depends on $\lambda_m$ as power law $\lambda
_{m}^{-\varepsilon /2} $. At low forces $\lambda _{m}$ is equal to $L$, so
that the renormalized (effective) friction coefficient scales as $f=w^{\ast
}\eta _{s}l^{d/2}L^{-\varepsilon /2}$. The drift velocity behaves
consequently as%
\begin{equation}
\mathrm{v}_{c}^{z}\simeq \frac{F}{fN}\simeq \frac{F}{\eta _{s}(Ll)^{d/2-1}},
\end{equation}%
which agrees in $d=3$ with the Stokes formula (\ref{v-Z}).

The inspection of the first-order correction to the longitudinal size of the
polymer yields that it is finite in four dimensions. This is what is
expected, because the friction coefficient, which is the only quantity to
renormalize, does not appear in the zero-order correction to the longitudinal size of
the polymer. Thus, the RG prediction for the longitudinal size consists in
replacing the bare expansion parameter in Eq.~(\ref{eq18}) by the
renormalized one. According to Eq.~(\ref{eq24}) the first-order correction is
of order $\varepsilon $ and is thus small in the vicinity of four dimensions.

For small forces the velocity and the elongation are given in the renormalized theory by expressions
\begin{equation}
\mathrm{v}_{c}^{z}=\frac{F}{fN}(1+\frac{1}{4}wB_{\mathrm{v}}(\beta F\sqrt{%
lL/2d})+\cdots),  \label{eq25}
\end{equation}
\begin{equation}
\left\langle r^{z}(0,t)\!-\!r^{z}(L,t)\right\rangle\!=\!\frac{FLl}{2dk_{B}T}%
(1-\frac{1}{4}wB_{\mathrm{r}}(\beta F\!\sqrt{lL/2d})+\cdots).  \label{eq26}
\end{equation}
The function $B_{\mathrm{v}}(y)$ in (\ref{eq25}) is defined by  $B_{\mathrm{v}}(y)=\int_{0}^{1}dx_{2}%
\int_{0}^{x_{2}}dx_{1}\frac{A(y)-A(0)}{(x_{2}-x_{1})^{d/2-1}}$.
The effective expansion parameter $w$ is a small number ($\sim{O(\epsilon)}$%
), so that the expansions (\ref{eq25}) and (\ref{eq26}) are reliable. The
expansion of the functions $B_{\mathrm{v}}(z)$ and $B_{r}(z)$ for small $z$
are given by%
\begin{eqnarray}
B_{\mathrm{v}}(z) &=&-\frac{1136}{14175}z^{2}+\frac{19366}{7882875}\label{Bvz}%
z^{4}+\cdots, \\
B_{r}(z) &=&\frac{64}{45}-\frac{464}{31185}z^{2}+\frac{10786}{19144125}\label{Brz}%
z^{4}+\cdots.
\end{eqnarray}
Expressions (\ref{eq25}-\ref{eq26}) demonstrate that the velocity and the longitudinal polymer size show the non-linear
response i.e. non Stokes behavior at moderate pulling forces.

Note that the Zimm result for the velocity (\ref{v-Z}) can be also obtained
in a simple way following the Kirkwood type approach to the diffusion
constant \cite{kirkwood}. That consists in neglecting one in the bracket of (\ref{eq13}). The
friction coefficient then drops, and the dependence on $N$ is the same as (%
\ref{v-Z}).

\subsection{Large elongation}

We now will study the first-order corrections \ to the velocity and the
longitudinal elongation for large argument $\beta F\sqrt{lL/2d}$. To
estimate the integrals in (\ref{eq13}, \ref{eq18}) for large forces we use
the asymptotic expression of $A(y)$ given by Eq.~(\ref{eq15}). To prevent the divergence in the integral over $%
t=x_{2}-x_{1}$ from the integration region $t=0$, we introduce an cutoff $%
t_{0}\simeq 6/Ll\beta ^{2}F^{2}$. The evaluation of the integrals over $t$
and $x_{2}$ can be performed analytically and gives for $d=3$%
\begin{equation}
\mathrm{v}_{c}^{z}=\frac{F}{f_0N}\left( 1+c_{\mathrm{v}}\xi _{0}L^{1/2}\frac{%
\ln (\beta F\left( Ll\right) ^{1/2})\ln \frac{\beta F\left( Ll\right) ^{1/2}%
}{3}}{\beta F\left( Ll\right) ^{1/2}}+\cdots \right)  \label{eq27}
\end{equation}%
with a numerical constant $c_{\mathrm{v}}$.
We would like to stress that the integral in Eq.~(\ref{eq13}) computed numerically can be fitted with $(\ln y)^{1.75}/y$ instead of the estimate given by Eq.~(\ref{eq27}). The difference is due to the complicated form of the expression for $y$ in Eq.~(\ref{def-y}).

For large $F$ and finite $L$ we
arrive at the Rouse result. However, for large $L$ and finite $F$ the
correction increases logarithmically with $L$, and will become large for
large $L$. Unfortunately, there are no analytical means to study the effect
of the whole perturbation expansion on $\mathrm{v}_{c}^{z}$ in this case. The
extrapolation of (\ref{eq27}) a la Kirkwood, i.e. the disregard of one in
the bracket of Eq.~(\ref{eq27}), yields%
\begin{equation}
\mathrm{v}_{c}^{z}\simeq \frac{k_BT}{\eta _{s}Nl^2}\ln (\beta F\left( Ll\right)
^{1/2})\ln \frac{\beta F\left( Ll\right) ^{1/2}}{3}.  \label{eq28}
\end{equation}%
Eq.~(\ref{eq28}) shows that in this regime (finite $F$ and large $L$) the
hydrodynamic interactions determine the behavior of the polymer. The friction coefficient drops in the expression of the velocity in this
regime. We expect that the hydrodynamic interactions remain important due to
the thermal fluctuations of the pulled chain. The presence of $k_BT$ in (\ref{eq28}) supports this statement. Note that the logarithmic
dependencies of $\mathrm{v}_{c}^{z}$ on the force does not allow to write
Eq.~(\ref{eq28}) in the form of Eq.~(\ref{v-R}) with some effective friction
coefficient. Eq.~(\ref{eq28}) can be formally obtained from Eq.~(\ref{v-R}) by
simultaneous replacements
\begin{equation}
f_{0}\rightarrow f\simeq l\eta _{s}/\ln (\beta F\left( Ll\right) ^{1/2}),\
F\rightarrow \frac{k_BT}{l}\ln \frac{\beta F\left( Ll\right) ^{1/2}}{3}.  \label{eq29}
\end{equation}%
The "renormalized" friction coefficient in Eq.~(\ref{eq29}) resembles to some
extent the friction coefficient, $L\eta _{s}/\ln (L/l)$, of a slender body (a cylinder of length $L$ and cross section radius $l$, $L \gg l$) in a flow \cite
{batchelor70,doi-edwards} (see also \cite{larson-chu97}).
However, the elongated polymer chain is different from a rod, and the comparison is only qualitativ.

A similar computation of the first-order correction to the chain size yields%
\begin{eqnarray}
&&\left\langle r^{z}(0,t)-r^{z}(L,t)\right\rangle =\frac{FLl}{2dk_{B}T}%
\left( 1-c_{r}\xi _{0}L^{1/2}\times \right.  \notag \\
&&\left. \frac{(\ln \frac{\beta F\left( Ll\right) ^{1/2}}{3}-2)\ln \left(
\beta F\left( Ll\right) ^{1/2}\right) }{\beta F\left( Ll\right) ^{1/2}}%
+\cdots\right) .  \label{eq30}
\end{eqnarray}
Note that the integral in Eq.~(\ref{eq18a}) computed numerically can be fitted with $(\ln y)^{2.47}/y$ instead of the analytical estimate in Eq.~(\ref{eq30}).
For large $F$ and finite $L$, we arrive at the Rouse result. On the contrary, for large $L$ and finite $F$ the first-order correction to the longitudinal size increases logarithmically with $L$. This results in a decrease of the longitudinal size of the polymer. It is difficult, due to the absence of an analytical method to make predictions on the total effect of the hydrodynamic interaction for polymer elongation.
Note that the corrections in Eqs.~(\ref{eq13}, \ref{eq18}) at $F\neq 0$
imply the nonlinear response of the polymer on the pulling force.

\section{Summary}
\label{concl}
To summarize, we have studied the drift of an ideal polymer driven by a constant force applied at
one polymer end using perturbation expansions in powers of hydrodynamic interactions. For moderate force, where the hydrodynamic interactions are not screened and the polymer elongation is small, the renormalization group method permits to compute the velocity and the longitudinal elongation of the polymer. These quantities are nonlinear functions of the driving force in this regime.
For large elongation we found two regimes. For large force but finite chain length $L$ the hydrodynamic interactions are screened, so that both the velocity and the longitudinal polymer elongation are given by the corresponding results of the Rouse theory. For large $L$ but finite force, the regime which we have studied for $d=3$, the hydrodynamic interactions are partially screened. The first-order corrections to $\mathrm{v}_{c}^{z}$ and $\left\langle r^{z}(0,t)\!-\!r^{z}(L,t)\right\rangle\nonumber$, increase logarithmically with $L$. Following the Kirkwood's treatment of the diffusion constant we make a prediction for the velocity, Eq.~(\ref{eq27}), beyond the first-order of the perturbation theory.
It would be of great interest to check experimentally these predictions by
pulling polymer in solvent using optical tweezers.

\begin{acknowledgments}
A financial support from the Deutsche Forschungsgemeinschaft, SFB 418, is
gratefully acknowledged.  We are grateful to Sriram Ramaswamy for useful comments. S.S. acknowledges also a partial financial support from the Deutsche Forschungsgemeinschaft, the grant Ste 981/3-1.
\end{acknowledgments}


\begin{thebibliography}{99}
\bibitem{doi-edwards} M. Doi and S. F. Edwards, The Theory of Polymer
Dynamics (Clarendon, Oxford, 1986).

\bibitem{kirkwood} J. Kirkwood, J. Polym. Sci. \textbf{21}, 1, 1954; J.
Kirkwood and J. J. Riseman, J. Chem. Phys. \textbf{16}, 565, 1948.

\bibitem{zimm} B. Zimm, J. Chem. Phys. \textbf{24}, 269, 1956.




\bibitem{larson-chu97} R.G. Larson, T. T. Perkins, D. E. Smith, and S. Chu,
Phys. Rev. E \textbf{55}, 1794 (1997).

\bibitem{perkins-smith-chu97} T. T. Perkins, D. E. Smith, and S. Chu,
Science \textbf{276}, 2016 (1997).

\bibitem{schoeder-chu03} C. M. Schroeder, H. P. Babcock, E. S. G. Shaqfeh,
and S. Chu, Science \textbf{301}, 1515 (2003).

\bibitem{brochard93} F. Brochard-Wyart, Europhys. Lett. \textbf{23}, 105 (1993).

\bibitem{zimmermann99} R. Rzehak, D. Kienle, T. Kawakatsu, and W. Zimmermann, Europhys. Lett. \textbf{46}, 821(1999).

\bibitem{zimmermann-epje00} R. Rzehak, W. Kromen, T. Kawakatsu, and W. Zimmermann, Eur. Phys. J. E \textbf{2}, 3 (2000).

\bibitem{freed} S. Q. Wang and K. F. Freed, J. Chem. Phys. \textbf{85}, 6210
(1986).

\bibitem{oono} Y. Oono, Adv. Chem. Phys. \textbf{61}, 301 (1985).

\bibitem{stepanow} S. Stepanow and G. Helmis, Phys. Rev. A \textbf{39}, 6037
(1989).

\bibitem{batchelor70} G. K. Batchelor, J. Fluid Mech. 44, 419 (1970).
\end{thebibliography}
\end{document}